\documentstyle[12pt]{article}

\catcode`\@=11
\def\@citex[#1]#2{%
\if@filesw \immediate \write \@auxout {\string \citation {#2}}\fi
\@tempcntb\m@ne \let\@h@ld\relax \def\@citea{}%
\@cite{%
  \@for \@citeb:=#2\do {%
    \@ifundefined {b@\@citeb}%
      {\@h@ld\@citea\@tempcntb\m@ne{\bf ?}%
      \@warning {Citation `\@citeb ' on page \thepage \space undefined}}%
      {\@tempcnta\@tempcntb \advance\@tempcnta\@ne%
      \@tempcntb\number\csname b@\@citeb \endcsname \relax%
      \ifnum\@tempcnta=\@tempcntb 
	\ifx\@h@ld\relax%
	  \edef \@h@ld{\@citea\csname b@\@citeb\endcsname}%
	\else%
	  \edef\@h@ld{\ifmmode{-}\else--\fi\csname b@\@citeb\endcsname}%
	\fi%
      \else
	\@h@ld\@citea\csname b@\@citeb \endcsname%
	\let\@h@ld\relax%
      \fi}%
    \def\@citea{,\penalty\@highpenalty\,}%
  }\@h@ld
}{#1}}

\def\@citeb#1#2{{[#1]\if@tempswa , #2\fi}}
\def\@citeu#1#2{{$^{#1}$\if@tempswa , #2\fi }}
\def\@citep#1#2{{#1\if@tempswa , #2\fi}}

\def\bcites{         
	\catcode`\@=11
	\let\@cite=\@citeb
	\catcode`\@=12
}

\def\upcites{         
	\catcode`\@=11
	\let\@cite=\@citeu
	\catcode`\@=12
}

\def\plaincites{      
	\catcode`\@=11
	\let\@cite=\@citep
	\catcode`\@=12
}

\newcount\hour
\newcount\minute
\newtoks\amorpm
\hour=\time\divide\hour by 60
\minute=\time{\multiply\hour by 60 \global\advance\minute by-\hour}
\edef\standardtime{{\ifnum\hour<12 \global\amorpm={am}%
	\else\global\amorpm={pm}\advance\hour by-12 \fi
	\ifnum\hour=0 \hour=12 \fi
	\number\hour:\ifnum\minute<10 0\fi\number\minute\the\amorpm}}
\edef\militarytime{\number\hour:\ifnum\minute<10 0\fi\number\minute}

\def\draftlabel#1{{\@bsphack\if@filesw {\let\thepage\relax
   \xdef\@gtempa{\write\@auxout{\string
      \newlabel{#1}{{\@currentlabel}{\thepage}}}}}\@gtempa
   \if@nobreak \ifvmode\nobreak\fi\fi\fi\@esphack}
	\gdef\@eqnlabel{#1}}
\def\@eqnlabel{}
\def\@vacuum{}
\def\marginnote#1{}
\def\draftmarginnote#1{\marginpar{\raggedright\scriptsize\tt#1}}
\def\draft{
	\pagestyle{plain}
	\overfullrule=2pt
	\oddsidemargin -.5truein
	\def\@oddhead{\sl \phantom{\today\quad\militarytime} \hfil
	\smash{\Large\sl DRAFT} \hfil \today\quad\militarytime}
	\let\@evenhead\@oddhead
	\let\label=\draftlabel
	\let\marginnote=\draftmarginnote
	\def\ps@empty{\let\@mkboth\@gobbletwo
	\def\@oddfoot{\hfil \smash{\Large\sl DRAFT} \hfil}
	\let\@evenfoot\@oddhead}
	\def\@eqnnum{(\theequation)\rlap{\kern\marginparsep\tt\@eqnlabel}%
	\global\let\@eqnlabel\@vacuum}  }

\def\section{\@startsection {section}{1}{\z@}{3.ex plus 1ex minus
 .2ex}{2.ex plus .2ex}{\large\bf}}
\def\subsection{\@startsection{subsection}{2}{\z@}{2.75ex plus 1ex minus
 .2ex}{1.5ex plus .2ex}{\bf}}

\def\appendix{{\newpage\section*{Appendices}}\let\appendix\section%
	{\setcounter{section}{0}
	\gdef\thesection{\Alph{section}}}\section}

\def\abstract{\if@twocolumn
\section*{Abstract}
\else 
\begin{center}
{\bf Abstract\vspace{-.5em}\vspace{0pt}}
\end{center}
\quotation
\fi}

\catcode`\@=12

\def\noj#1,#2,{{\bf #1} (19#2)\ }
\def\jou#1,#2,#3,{{\sl #1\/ }{\bf #2} (19#3)\ }
\def\ann#1,#2,{{\sl Ann.\ Physics\/ }{\bf #1} (19#2)\ }
\def\cmp#1,#2,{{\sl Comm.\ Math.\ Phys.\/ }{\bf #1} (19#2)\ }
\def\cq#1,#2,{{\sl Class.\ Quantum Grav.\/ }{\bf #1} (19#2)\ }
\def\cqg#1,#2,{{\sl Class.\ Quantum Grav.\/ }{\bf #1} (19#2)\ }
\def\ijmp#1,#2,{{\sl Int.\ J.\ Mod.\ Phys.\/ }{\bf A#1} (19#2)\ }
\def\jmp#1,#2,{{\sl J.\ Math.\ Phys.\/ }{\bf #1} (19#2)\ }
\def\lmp#1,#2,{{\sl Lett.\ Math.\ Phys.\/ }{\bf #1} (19#2)\ }
\def\grg#1,#2,{{\sl Gen.\ Rel.\ Grav.\/ }{\bf #1} (19#2)\ }
\def\mpl#1,#2,{{\sl Mod.\ Phys.\ Lett.\/ }{\bf A#1} (19#2)\ }
\def\nc#1,#2,{{\sl Nuovo Cim.\/ }{\bf #1} (19#2)\ }
\def\np#1,#2,{{\sl Nucl.\ Phys.\/ }{\bf B#1} (19#2)\ }
\def\pl#1,#2,{{\sl Phys.\ Lett.\/ }{\bf #1B} (19#2)\ }
\def\pla#1,#2,{{\sl Phys.\ Lett.\/ }{\bf #1A} (19#2)\ }
\def\pr#1,#2,{{\sl Phys.\ Rev.\/ }{\bf #1} (19#2)\ }
\def\prd#1,#2,{{\sl Phys.\ Rev.\/ }{\bf D#1} (19#2)\ }
\def\prl#1,#2,{{\sl Phys.\ Rev.\ Lett.\/ }{\bf #1} (19#2)\ }
\def\prp#1,#2,{{\sl Phys.\ Rept.\/ }{\bf #1C} (19#2)\ }
\def\ptp#1,#2,{{\sl Prog.\ Theor.\ Phys.\/ }{\bf #1} (19#2)\ }
\def\ptpsup#1,#2,{{\sl Prog.\ Theor.\ Phys.\/ Suppl.\/ }{\bf #1} (19#2)\ }
\def\rmp#1,#2,{{\sl Rev.\ Mod.\ Phys.\/ }{\bf #1} (19#2)\ }
\def\yadfiz#1,#2,#3[#4,#5]{{\sl Yad.\ Fiz.\/ }{\bf #1} (19#2) #3%
\ [{\sl Sov.\ J.\ Nucl.\ Phys.\/ }{\bf #4} (19#2) #5]}
\def\zh#1,#2,#3[#4,#5]{{\sl Zh.\ Exp.\ Theor.\ Fiz.\/ }{\bf #1} (19#2) #3%
\ [{\sl Sov.\ Phys.\ JETP\/ }{\bf #4} (19#2) #5]}

\def\beq{\begin{equation}}
\def\eeq{\end{equation}}
\def\beqar{\begin{eqnarray}}
\def\eeqar{\end{eqnarray}}

\def\nfrac#1#2{{\displaystyle{\vphantom1\smash{\lower.5ex\hbox{\small$#1$}}%
	\over\vphantom1\smash{\raise.25ex\hbox{\small$#2$}}}}}

\def\p#1{\mskip#1mu}

\def\stop{\p6.}
\def\comma{\p6,}

\def\pa{\partial}

\def\l:{\mathopen{:}\,}
\def\r:{\,\mathclose{:}}

\def\[{\left[}          \def\]{\right]}
\def\({\left(}          \def\){\right)}
\def\<{\left<}          \def\>{\right>}

\def\CT{{\cal T}}
\def\CW{{\cal W}}
\def\CY{{\cal Y}}

\def\la{\langle}
\def\ra{\rangle}

\begin{document}
\begin{titlepage}

\begin{center}
Oct 16, 1994\hfill      
\hfill                  WIS-94/45/Oct-PH               \\
\hfill                  hep-th/9410157

\vskip 1 cm
{\large \bf c=1 discrete states correlators via W$_{1+\infty}$ constraints\\}
\vskip 0.1 cm
\vskip 0.5 cm
{Amihay Hanany
}
\vskip 0.2cm
{\sl
ftami@wicc.weizmann.ac.il \\
Department of Particle Physics \\
Weizmann Institute of Science \\
 76100 Rehovot Israel
}
\vskip 0.2cm
{Yaron Oz\footnote{Work
supported in part by the US-Israel Binational Science Foundation,
and the Israel Academy of Science.
}
}
\vskip 0.2cm

{\sl
yarono@ccsg.tau.ac.il \\
School of Physics and Astronomy\\Raymond and Beverly Sackler Faculty
of Exact Sciences\\Tel-Aviv University\\Ramat Aviv, Tel-Aviv 69978, Israel.
}

\end{center}

\vskip 0.5 cm
\begin{abstract}
The discrete states of $c=1$ string theory at the self-dual radius
are associated with modes of
$W_{1+\infty}$ currents and their genus zero correlators are computed.
An analogy to a recent suggestion based on
the integrable structure of the theory is found.
An iterative method for deriving the dependence of the currents on the
full space of couplings is presented and applied.
The dilaton equation of the theory is derived.

\end{abstract}

\end{titlepage}

The discrete states of $c=1$ string theory have been originally
found in the Liouville
formulation \cite{Lian,Mukherji,Bouk,Wittds,PolyK} and interpreted as $2d$
 remnants of transverse massive
string excitations \cite{Polys,Polyakov}.
The computation of their correlators in the presence
of the cosmological constant in this framework is rather non-trivial
and has not been carried out successfully yet.

The interpretation of the discrete states as gravitational descendants
in the
topological description of $2d$ string theory
at the self-dual radius \cite{HOP,GM} opened a way
to the computation of the correlators by other means.
In \cite{LOS} the correlators have been computed by using topological
recursion relations derived via analytical continuation of those
of the minimal models.

Discrete states have been introduced in the matrix model approach by
associating them with powers of the eigenvalue matrix \cite{Dan,BXd}.
Their correlators have been calculated in this formulation with results
that do not coincide with those of \cite{LOS}.

In this letter we take another route for defining the discrete states.
In analogy with $(1,q)$ minimal topological models in which gravitational
descendants are associated with modes of the $W_q$ currents
\cite{DVV,FKN,Goeree}, we
associate the discrete states with modes of the $W_{1+\infty}$ constraints
 algebra
of $c=1$ string theory \cite{DMP} and compute their correlators.
We find analogy to the suggestion made in \cite{Takasaki} based on the
integrable
structure of the theory.
The results that we get for the discrete states correlators
do not agree neither with \cite{LOS} nor with \cite{Dan,BXd}, and the
implications of this will be discussed at the end of the paper.

The $W$ constraints on the partition function of minimal topological
matter coupled to topological gravity read \cite{DVV,FKN,Goeree}.
\beq
{\mu}^{-2}Z^{-1}\partial_{k,\alpha}Z=
\langle\langle\sigma_{k,\alpha}\rangle\rangle =
Z^{-1}W_{k-\alpha}^{(\alpha+1)}Z
 \stop
\label{W}
\eeq
$\sigma_{k,\alpha}$ is the $k^{th}$ gravitational descendant of
the primary field $O_{\alpha}$ and $W_m^{(\alpha+1)}$ is the $m^{th}$
mode of the spin $\alpha+1$ current.
The partition function is:
\beq
\log Z(t) = \mu^2 \langle exp(\sum_{k,\alpha}t_{k,\alpha}\sigma_{k,\alpha})
\rangle \comma
\eeq
and
\beq
\langle\cdots\rangle \equiv \sum_{g \geq 0}
\frac{1}{\mu^{2g}}\langle\cdots\rangle_g  \comma
\eeq
with $g$ being the genus of the Riemann surface.

The $W_{1+\infty}$ Ward identities for the tachyon correlators in
$2d$ string theory read \cite{DMP}
\beq
\langle\langle \CT_n\rangle\rangle = Z^{-1}\bar{W}_{-n}^{(n+1)}Z \comma
\label{tWI}
\eeq
where $\bar{W}^{(n+1)}$ is the spin $n+1$ current of a $W_{1+\infty}$ algebra
and
\beq
\langle\langle \CT_n\rangle\rangle = \langle \CT_n
 exp(\sum_{k=-\infty}^{\infty}t_k\CT_k)\rangle \stop
\eeq
By the tachyon $\CT_n$ we mean the Seiberg state $\CT_n^+$ with
positive momentum $n$ \cite{Seiberg}.
Analogous $W_{1+\infty}$ Ward identities exist for a
negative momentum tachyon
$\CT_{-n}^+$
\beq
\langle\langle \CT_{-n}\rangle\rangle = Z^{-1}W_{-n}^{(n+1)}Z \comma
\label{tWI2}
\eeq
where $W^{(n+1)}$ is the spin $n+1$ current of a similar $W_{1+\infty}$
algebra.

Comparing the Ward identities (\ref{W}) for the primary operators
$\sigma_{0,\alpha}$ and (\ref{tWI}) we are led to
identify $\CT_n$ as primaries in the topological description of $2d$ string
theory.
This identification has been made in \cite{HOP} using integrable and
topological reasoning.
In analogy with (\ref{W}), we may try to generalize
(\ref{tWI}) by using the other
modes of the $W_{1+\infty}$ currents, namely
\beq
\langle\langle\sigma_k(\CT_n)\rangle\rangle = Z^{-1}\bar{W}_{k-n}^{(n+1)}Z
\stop
\label{disWI}
\eeq
Upon identifying the discrete states as gravitational descendants
in the topological formulation of the theory \cite{HOP,GM}
\beq
{\cal Y}_{J,m} = \sigma_k(\CT_n) \comma
\eeq
with $k = J-m, n = J + m$,
we have
\beq
\langle\langle {\cal Y}_{J,m}\rangle\rangle = Z^{-1}\bar{W}_{-2m}^{(J+m+1)}Z
 \comma
\label{yWI}
\eeq
with $J$ taking half integer values and $-J \leq m \leq J$.

The currents in (\ref{W}) depend
on the times associated with both primaries and descendants, i.e.\ they
are defined on the full phase space.
Thus, we will assume in the paper that equation (\ref{yWI}) holds
on the full phase space too.
 From the $W_{1+\infty}$ Ward identities we only know
the dependence of the currents in (\ref{yWI}) on the times
associated with the tachyons.
As we will show in the sequel we can in fact find the dependence
of the currents in (\ref{yWI}) on the times associated with the
discrete states, thus enlarging the $W_{1+\infty}$ Ward identities to
the full phase space
\beq
\langle\langle {\cal Y}_{J,m}\rangle\rangle = Z^{-1}
\bar{{\cal W}}_{-2m}^{(J+m+1)}Z
 \comma
\label{yWI2}
\eeq
where on the space of only tachyon times $t_n$,
$\bar{{\cal W}}_{-2m}^{(J+m+1)}(t_n) \equiv \bar{W}_{-2m}^{(J+m+1)}(t_n)$.

In the following we will consider the genus zero case.
Equation (\ref{yWI}) on the space of tachyon times reads \cite{DMP}
\beq
\langle\langle {\cal Y}_{J,m}\rangle\rangle_0 =
\frac{1}{J+m+1}\oint x^{J-m}
\bar{W}^{J+m+1} \comma
\label{syWI}
\eeq
where
\beq
\bar{W}  = \frac{1}{x}[1 + \sum_{k>0}k t_{-k}x^k +
\sum_{k>0}x^{-k}
\langle\langle \CT_{-k}\rangle\rangle_0]  \stop
\label{barW}
\eeq

A priori, equation (\ref{syWI}) should be used only for positive momentum
discrete states ${\cal Y}_{J,m}, m>0$, while for negative momenta its
parity transformed version
\beq
\langle\langle {\cal Y}_{J,-m}\rangle\rangle_0 =
\frac{1}{J+m+1}\oint x^{J-m}
W^{J+m+1} \comma
\label{syWI2}
\eeq
where
\beq
W  = \frac{1}{x}[1 + \sum_{k>0}k t_kx^k +
\sum_{k>0}x^{-k}
\langle\langle \CT_k\rangle\rangle_0]  \comma
\label{barW2}
\eeq
should be used.
The $c=1$ string equation \cite{Takasaki,EK},
which can be written in the form \cite{HOP}
\beq
x = \bar{W}(W(x)) \comma
\label{seq}
\eeq
leads to the identity\footnote{The identity (\ref{id}) is a special case
of a more general one : $\oint f(x)G(W) = \oint g(x)F(\bar{W})$, where
$f(x),g(x)$ are any two functions and $F'(x) = f(x), G'(x)=g(x)$.}
\beq
\frac{1}{J+m+1}\oint x^{J-m}\bar{W}^{J+m+1}
=\frac{1}{J-m+1}\oint x^{J+m}W^{J-m+1}
\comma
\label{id}
\eeq
and therefore
we can use (\ref{yWI}) for negative values of $m$ as well.

It is interesting to notice that equation (\ref{syWI})
coincides with the suggestion of \cite{Takasaki} for discrete
states correlators. The latter was based on studying the
symmetries of the Toda lattice hierarchy in the dispersionless limit.

Equation (\ref{yWI2}) provides a complete and self-contained
definition of all the tachyon and discrete states correlators.
Equation (\ref{syWI}) is valid on the full phase space, with
$\bar{W}(t_n)$ being replaced by $\bar{{\cal W}}(t_{J,m})$
which we have to construct.
Note that we implicitly assume that the string equation (\ref{seq}) holds
on the full phase space too
\beq
x = \bar{{\cal W}}({\cal W}(x)) \stop
\label{seq2}
\eeq

Equation (\ref{syWI}) for the negative momentum tachyon $\CT_{-n = -2J}
\equiv {\cal Y}_{J,-J}$ on the full phase space reads:
\beq
\langle\langle \CT_{-n} \rangle\rangle_0 = \oint
x^n \bar{{\cal W}}
\stop
\eeq
This implies that
\beq
\bar{{\cal W}} = \frac{1}{x}[1 + \sum_{k>0}x^k
U_k({t_{J,m}})  +
\sum_{k>0}x^{-k}
\langle\langle \CT_{-k}\rangle\rangle_0] \comma
\label{barWfps}
\eeq
where $\langle\langle \cdots \rangle\rangle$ is defined on the full phase
space, and $U_k$ are unknown functions.
It is tempting to claim,
using (\ref{syWI}) with $J \rightarrow -J$,
that the functions $U_k$ in
(\ref{barWfps}) are given by
\beq
U_k({t_{J,m}}) = \langle\langle {\cal Y}_{-J,J} \rangle\rangle_0 \equiv
\langle\langle \CT_{k=2J}^- \rangle\rangle_0 \comma
\label{U}
\eeq
where $\CT_k^- $ is an anti-Seiberg state.
This is incorrect, since due to (\ref{barW}), $U_k=kt_{-k}$
on the space of tachyon times. This
together with (\ref{U}) implies the
vanishing of the correlators
$\langle \CT_n^- \prod_{i=1}^s\CT_{n_i}^+ \rangle = 0$
for $s > 2$ which is in contradiction with calculations of bulk correlators
in the Liouville formulation \cite{Polys,Polyakov,KdF}.
$U_k$ can be determined perturbatively in $t_{J,m}$ as we will show
now.

Consider first correlators with insertion of
one discrete state and a few tachyons by using
\beq
\langle{\cal Y}_{J,m}\prod_{i=1}^l\CT_{n_i}\rangle_g =
\pa_{n_1}\cdots\pa_{n_l}\langle\langle{\cal Y}_{J,m}\rangle\rangle_g
(t=0) \stop
\eeq
Using (\ref{syWI}),(\ref{barW}) and (\ref{syWI2}),(\ref{barW2}) we calculate
the
 first few genus
 zero
correlators:
\beqar
\langle{\cal Y}_{J,m}\rangle_0 & = & \frac{1}{J+1}\delta_{m,0} \nonumber\\
\langle{\cal Y}_{J,m}\CT_n\rangle_0 & = & 2\vert m\vert\delta_{n+2m,0}
 \nonumber\\
\langle{\cal Y}_{J,m}\CT_{n_1}\CT_{n_2}\rangle_0 & = &
(J+|m|)|n_1n_2|\delta_{n_1+n_2+2m,0} \nonumber\\
\langle{\cal Y}_{J,m}\prod_{i=1}^3\CT_{n_i}\rangle_0 & = &
(J+|m|)\prod_{i=1}^3|n_i|
(J-|m|+2m-1 -\sum_{j=1}^3(2m+n_j)\nonumber\\
&   &\theta(-2m -n_j))
\delta_{n_1+n_2+n_3+2m,0}.
\label{YTT}
\eeqar

The first three correlators in (\ref{YTT}) are explicitly parity invariant,
$(m \leftrightarrow -m, n_i \leftrightarrow -n_i)$,
 while the four point function
is parity invariant as a consequence of the identity
$2m-\sum_{j=1}^3(2m+n_j)\theta(-2m -n_j)={1\over2}\sum_{i=1}^3|2m+n_i|$.
The dependence of the $W_{1+\infty}$ constraints on the times associated
with the discrete states can be found iteratively as follows:
Consider
\beq
\langle{\cal Y}_{J,-m}\CT_n\rangle_0  =  2m\delta_{n,2m} =
\pa_{J,-m}\langle\langle \CT_n\rangle\rangle_0(t=0)  \comma
\label{YT}
\eeq
where $\pa_{J,m} \equiv \frac{\pa}{\pa t_{J,m}}$.
It implies that we have to modify $\bar{W}$ by adding to it
the term
\beq
\sum_{J;0<m\leq J}2m t_{J,-m}x^{2m-1} \stop
\eeq
For $J=m$ it coincides with the corresponding term for the tachyons.
Using
\beq
\langle{\cal Y}_{J,m}\CT_{-n}\rangle_0  =  2m\delta_{n,2m} =
\pa_{J,m}\langle\langle \CT_{-n}\rangle\rangle_0(t=0)  \comma
\label{nYT}
\eeq
we see that
we have to add a similar term of
the form
\beq
\sum_{J;0<m\leq J}2mt_{J,m}x^{-2m-1} \stop
\eeq
Thus,
\beq
\bar{{\cal W}} = \frac{1}{x}[1 + \sum_{J;-J \leq m \leq J}
2|m| t_{J,m}x^{-2m} + O(t^2)] \stop
\label{barWt}
\eeq
This can be used in order to compute the two-point function of discrete
states
\beq
\langle{\cal Y}_{J_1,m}{\cal Y}_{J_2,-m}\rangle_0  =
\pa_{J_1,m}\langle\langle{\cal Y}_{J_2,-m}\rangle\rangle_0(t=0) = 2|m| \stop
\label{YY}
\eeq

Consider now the correlator $\langle{\cal Y}_{J,m}\CT_{n_1}\CT_{n_2}\rangle_0$.
We can use it in order to get the next correction to $\bar{\CW}$ that
takes the form
\beq
\sum_{J;m,n}C_{J,m,n}t_{J,m}t_nx^{-2m-n-1} \comma
\label{tt}
\eeq
where
\beq
C_{J,m,n} = |n||n+2m|[(J+|m|)\theta(2m+n) +
(J-|m|)\theta(-2m-n)]\stop
\label{cJmn}
\eeq
Using (\ref{cJmn}) we can calculate the three point function
\beq
\langle{\cal Y}_{J_1,m_1}{\cal Y}_{J_2,m_2}\CT_n\rangle_0  =
4(J_1|m_2|+J_2|m_1|)|m_1+m_2|\delta_{2m_1+2m_2+n,0} \stop
\label{YYT}
\eeq

 From (\ref{YYT}) we deduce the full order $(t^2)$ term in
$\bar{{\cal W}}$,
it reads
\beq
\sum_{J_1,J_2; m_1,m_2}C_{J_1,m_1,J_2,m_2}
t_{J_1,m_1}t_{J_2,m_2}x^{-2m_1-2m_2-1} \comma
\eeq
with
\beq
C_{J_1,m_1,J_2,m_2} = 2(J_1|m_2|+J_2|m_1|)|m_1+m_2|+4|m_1m_2|(m_1+m_2)
\theta(-m_1-m_2)\stop
\label{Cjm12}
\eeq
The three point function of discrete states is thus calculated to be
\beq
\langle{\cal Y}_{J_1,m_1}{\cal Y}_{J_2,m_2}{\cal Y}_{J_3,m_3}\rangle_0  =
4(J_1|m_2m_3| + J_2|m_1m_3| + J_3|m_1m_2| - |m_1m_2m_3|)
\comma
\label{YYY}
\eeq
where momentum conservation implies $m_1+m_2+m_3 = 0$.
When one of the discrete states is a tachyon, equation (\ref{YYY})
reduces to (\ref{YYT}), while if two of them are tachyons it reduces
to (\ref{YTT}).
Note that formally
\beq
C_{J_1,m_1,J_2,m_2} =
{1\over2}\langle{\cal Y}_{J_1,m_1}{\cal Y}_{J_2,m_2}{\cal Y}_{m_1+m_2,-m_1-m_2}
\rangle_0
\comma
\label{Cjm12id}
\eeq
as indeed follows from the definition (\ref{syWI}).
Such relations hold for higher point functions too, and are useful for
extracting the structure of $\bar{{\cal W}}$ from the discrete states
correlators.

The same procedure leads to the four point function of discrete states
\beqar
\langle{\cal Y}_{J_1,m_1}{\cal Y}_{J_2,m_2}{\cal Y}_{J_3,m_3}
{\cal Y}_{J_4,m_4}\rangle_0 &=&
 8|m_1m_2m_3|[J_4^2-m_4^2-J_4(2 \max \{|m_i|\}+1)]+ perm  \nonumber \\
 &+& 8J_1J_2|m_3m_4|(2 \max \{|m_i|\}-|m_1+m_2|) + perm  \nonumber \\
 &+& 16|m_1m_2m_3m_4|(2 \max \{|m_i|\}+1)
\label{YYYY}
\eeqar
When three of the operators are tachyons the correlation function (\ref{YYYY})
reduces to the equation (\ref{YTT}).
The full order $(t^3)$ term in $\bar{{\cal W}}$ reads
\beq
\sum_{J_1,J_2,J_3; m_1,m_2,m_3}C_{J_1,m_1,J_2,m_2,J_3,m_3}
t_{J_1,m_1}t_{J_2,m_2}t_{J_3,m_3}x^{-2m_1-2m_2-2m_3-1} \comma
\eeq
with
\beq
C_{J_1,m_1,J_2,m_2,J_3,m_3} = {1\over6}
\langle\CY_{J_1,m_1}\CY_{J_2,m_2}\CY_{J_3,m_3}\CY_{m_1+m_2+m_3,-m_1-m_2-m_3}
\rangle_0
\stop
\label{Cjm123id}
\eeq
The iterative procedure described above can be used to find the
complete dependence of the $W_{1+\infty}$ currents on the full phase space.

The description of $c=1$ string theory at the self-dual radius as a
topological field theory should include the puncture and dilaton equations.
It has been observed that the momentum one tachyon corresponds to the
puncture operator \cite{kita,HOP}, thus the ${\cal T}_1$ Ward identity
is the puncture equation. We will now derive the dilaton equation.
The dilaton is the first descendant of the puncture. Thus, consider
\beq
\langle\langle {\cal Y}_{1,0}\rangle\rangle = Z^{-1}\bar{W}_0^{(2)}Z
 \comma
\label{Y10d}
\eeq
where $\bar{W}_0^{(2)}$ is the zero mode of the Virasoro current.
Equation (\ref{Y10d}) reads to all genera \cite{DMP}
\beqar
\la\la{\cal Y}_{1,0}\ra\ra =
\frac{1}{2}\oint \frac{1}{x}
[1 + \sum_{k>0}k t_{-k}x^k +
\sum_{k>0}x^{-k}
\langle\langle \CT_{-k}\rangle\rangle]^2 =  \nonumber\\
= {1\over2}+\sum_{g;k > 0}\frac{1}{\mu^{2g}}
kt_{-k}\la\la \CT_{-k} \ra\ra_g  \stop
\label{Y10id}
\eeqar

It yields
\beq
\langle{\cal Y}_{1,0}\prod_{i=1}^m\CT_{n_i}\rangle_g = \pa_{n_1}\cdots\pa_{n_m}
\la\la {\cal Y}_{1,0}\ra\ra_g (t=0) =
(\frac{1}{2}\sum_{i=1}^m|n_i|)\la\prod_{j=1}^m\CT_{n_j}\ra_g
\stop
\label{Y10}
\eeq

Using the $\CT_0$ Ward identity\footnote{This identity can be derived,
for instance in the continuum, by shifting the constant mode of the Liouville
 field.}
\beq
\langle \CT_0 \prod_{i=1}^m\CT_{n_i}\rangle_g =
(\frac{1}{2}\sum_{i=1}^m|n_i| + 2 -2g - m)\la\prod_{j=1}^m\CT_{n_j}\ra_g
\comma
\label{T0}
\eeq
with (\ref{Y10}), we see that the operator ${\cal D} \equiv \CT_0
-{\cal Y}_{1,0}$ satisfies
\beq
\langle{\cal D} \prod_{i=1}^m\CT_{n_i}\rangle_g =
(2 -2g - m)\la\prod_{j=1}^m\CT_{n_j}\ra_g
\stop
\label{D}
\eeq
Equation (\ref{D}) is the dilaton equation and ${\cal D}$ is
the dilaton operator which measures the Euler characteristic of
the Riemann surface with $m$ punctures, that is $2-2g-m$.

There are several questions that arise as a consequence of this work.
We presented an iterative method for computing the
constraints currents on the full phase space and
carried the computation to a certain order in $t_{J,m}$.
It would be interesting
to find the complete constraints algebra on the full phase space and
for arbitrary genus.
This algebra should have a topological interpretation along the lines of
\cite{LOS} or \cite{GIM}, and an underlying integrable hierarchy generalizing
the Toda lattice, that should be discovered.
For that the integrable viewpoint of \cite{Takasaki,Nakatsu} is likely to be
 helpful.
Furthermore, one expects these results to be intimately related to
intersection theory on the moduli space of Riemann surfaces.

The results for discrete states correlators as calculated in this letter
do not coincide with those calculated in \cite{LOS,BXd}.
This implies that there is more than one way to perturb the tachyon theory,
by introducing extra operators. Thus, the main, unanswered yet, question
is which of the various theories is equivalent to the
Liouville $c=1$ string theory.

\end{document}